\documentclass[aps,amsmath,amssymb,superscriptaddress,twocolumn,nofootinbib,showpacs,preprintnumbers,notitlepage,prl]{revtex4-2}

\usepackage{amsmath}
\usepackage{amsfonts}
\usepackage{amssymb}
\usepackage{physics}
\usepackage{slashed}
\usepackage{simplewick}
\usepackage{mathtools}
\usepackage{braket}
\usepackage{bm} 
\usepackage{dsfont}

\usepackage{siunitx}
\usepackage{orcidlink}
\usepackage{placeins}
\usepackage{graphicx}
\usepackage{float}
\usepackage[caption=false]{subfig}

\usepackage{listings}

\usepackage{hyperref}
\hypersetup{ 
    pdfnewwindow=true,      
    colorlinks=true,       
    allcolors=[RGB]{31 119 180}
} 
\usepackage{tikz}
\usetikzlibrary{calc}
\usetikzlibrary{patterns}
\usetikzlibrary{patterns}
\usetikzlibrary{positioning,decorations.pathmorphing,decorations.markings,snakes}

\usepackage{xcolor}
\definecolor{jpac-blue}{rgb}{0.12,0.47,0.71}
\definecolor{jpac-orange}{rgb}{1,0.5,0.05}
\definecolor{jpac-green}{rgb}{0,0.62,0.45}
\definecolor{jpac-red}{rgb}{0.84,0.15,0.15}
\definecolor{jpac-purple}{rgb}{0.58,0.40,0.71}
\definecolor{jpac-brown}{rgb}{0.54,0.34,0.29}
\definecolor{jpac-pink}{rgb}{0.89,0.47,0.76}
\definecolor{jpac-grey}{rgb}{0.5,0.5,0.5}
\definecolor{jpac-gold}{rgb}{0.74,0.74,0.13}
\definecolor{jpac-aqua}{rgb}{0.09,0.75,0.81}
\definecolor{jpac-white}{rgb}{1,1,1}

\usepackage[capitalise]{cleveref}

\usepackage{mathrsfs}
\usepackage{pgfplots,pgfplotstable}
\usetikzlibrary{pgfplots.groupplots}
\usetikzlibrary{decorations.markings}
\usepgfplotslibrary{fillbetween}
\pgfplotsset{compat=1.14}
\usepackage{tikz}
\usepackage{tikz-3dplot}
\usetikzlibrary{shapes.geometric, arrows.meta, positioning}
\usetikzlibrary{calc}
\tikzset{->-/.style={decoration={
  markings,
  mark=at position .5 with {\arrow{>}}},postaction={decorate}}}
\tikzset{>=stealth}
\makeatletter
\tikzoption{canvas is plane}[]{\@setOxy#1}
\def\@setOxy O(#1,#2,#3)x(#4,#5,#6)y(#7,#8,#9)%
  {\def\tikz@plane@origin{\pgfpointxyz{#1}{#2}{#3}}%
   \def\tikz@plane@x{\pgfpointxyz{#4}{#5}{#6}}%
   \def\tikz@plane@y{\pgfpointxyz{#7}{#8}{#9}}%
   \tikz@canvas@is@plane
  }
\makeatother 

\newcommand\bsub{\begin{subequations}}
\newcommand\esub{\end{subequations}}

\usepackage{multirow}
\usepackage{dcolumn}
\usepackage{tabularx}

\usepackage{array}   
\newcolumntype{L}{>{$}l<{$}} 
\newcolumntype{R}{>{$}r<{$}}
\newcolumntype{C}{>{$}c<{$}}
\usepackage{xspace}
\usepackage[normalem]{ulem}
\usepackage[format=plain,justification=raggedright,singlelinecheck=false]{caption}
\usepackage[format=plain,justification=raggedright,singlelinecheck=false]{subcaption}
\usepackage{nameref}

\newcommand{\mevnospace}{\ensuremath{{\mathrm{\,Me\kern -0.1em V}}}}
\newcommand{\gevnospace}{\ensuremath{{\mathrm{\,Ge\kern -0.1em V}}}}
\newcommand{\tevnospace}{\ensuremath{{\mathrm{\,Te\kern -0.1em V}}}}

\newcommand{\eg}{{\it e.g.}\xspace}

\newcommand{\cf}{{\it cf.}\xspace}

\newcommand{\comment}[1]{}

\usepackage{diagbox}
\usepackage{todonotes}
\usepackage{enumitem}

\newcommand{\mytitle}[1]{\vspace{.5cm}{\em #1.---}}

\newcommand{\DelBar}{\Delta(1232)}

\newcommand{\helgamma}{\lambda_\gamma}
\newcommand{\helDel}{\lambda_\Delta}
\newcommand{\helproton}{\lambda_N}
\newcommand{\helgammat}{\mu_\gamma}
\newcommand{\helDelt}{\mu_\Delta}
\newcommand{\helprotont}{\mu_{\bar{N}}}

\newcommand{\AGH}{AGH University of Krakow, Faculty of Physics and Applied Computer Science, PL-30-059 Krak\'ow, Poland}
\newcommand{\catania}{INFN Sezione di Catania, I-95123 Catania, Italy}
\newcommand{\ceem}{Center for  Exploration  of  Energy  and  Matter, Indiana  University, Bloomington,  IN  47403,  USA}
\newcommand{\indiana}{Department of Physics, Indiana  University, Bloomington,  IN  47405,  USA}
\newcommand{\jlab}{Theory Center, Thomas  Jefferson  National  Accelerator  Facility, Newport  News,  VA  23606,  USA}
\newcommand{\messina}{Dipartimento di Scienze Matematiche e Informatiche, Scienze Fisiche e Scienze della Terra, Universit\`a degli Studi di Messina, I-98122 Messina, Italy}
\newcommand{\ub}{Departament de F\'isica Qu\`antica i Astrof\'isica and Institut de Ci\`encies del Cosmos, Universitat de Barcelona, E-08028 Barcelona, Spain}
\newcommand{\uned}{Departamento de F\'isica Interdisciplinar, Universidad Nacional de Educaci\'on a Distancia (UNED), E-28040 Madrid, Spain}

\newcommand{\odu}{Department of Physics, Old Dominion University, Norfolk, VA 23529, USA}
\newcommand{\MIT}{Center for Theoretical Physics - A Leinweber Institute, Massachusetts Institute of Technology, Cambridge, MA 02139, USA}
\newcommand{\icn}{Instituto de Ciencias Nucleares,
Universidad Nacional Aut\'onoma de M\'exico, Ciudad de M\'exico 04510, Mexico}

\begin{document}
\preprint{JLAB-THY-26-4778}
\preprint{MIT-CTP/6046}

\title{First determination of vector and tensor couplings from polarized $\pi\Delta$ photoproduction}

\author{Vanamali~\surname{Shastry}\orcidlink{0000-0003-1296-8468}}
\email{vanamalishastry@gmail.com}
\affiliation{\ceem}
\affiliation{\indiana}
\author{{\L}ukasz Bibrzycki\orcidlink{0000-0002-6117-4894}}
\affiliation{\AGH}
\author{Vincent~Mathieu\orcidlink{0000-0003-4955-3311}}
\affiliation{\ub}
\author{Gl\`oria~\surname{Monta\~na}\orcidlink{0000-0001-8093-6682}}
\affiliation{\ub}
\author{Alessandro~\surname{Pilloni}\orcidlink{0000-0003-4257-0928}}
\affiliation{\messina}
\affiliation{\catania}
\author{C\'esar~\surname{Fern\'andez-Ram\'irez}\orcidlink{0000-0001-8979-5660}}
\affiliation{\uned}
\author{Robert~J.~\surname{Perry}\orcidlink{0000-0002-2954-5050}}
\affiliation{\MIT}
\author{Arkaitz~\surname{Rodas}\orcidlink{0000-0003-2702-5286}}
\affiliation{\jlab}
\affiliation{\odu}
\author{Adam~P.~\surname{Szczepaniak}\orcidlink{0000-0002-4156-5492}}
\affiliation{\ceem}
\affiliation{\indiana}
\affiliation{\jlab}
\author{Daniel~\surname{Winney}\orcidlink{0000-0002-8076-243X}}
\affiliation{\icn}
\collaboration{Joint Physics Analysis Center}

%
%
\begin{abstract}
    The couplings between hadrons encode the dynamics of quantum chromodynamics. While many couplings can be calculated from decay widths, in some cases the decays are kinematically forbidden and hence are not directly accessible. We use a Regge framework to determine these couplings from high-energy polarized scattering processes. We apply this to the $\pi\Delta$ photoproduction that was recently studied at GlueX and provide the first determination of the complete set of $N\Delta$ couplings to $\rho$, $b_1$, and $a_2$.
\end{abstract}

\maketitle

\mytitle{Introduction} Understanding the interactions between hadrons requires the determination of the strengths of their couplings to various channels, which encode the underlying dynamics of QCD. Among the various interactions, the couplings of the $N\Delta$ system to light mesons are of particular interest because of their applications to a wide range of problems. The role of $\Delta$ as an intermediate state in various processes involving the interactions of nucleons and pions has been studied extensively~\cite{Hemmert:1997ye,Pascalutsa:1998pw,Pascalutsa:2000kd}. The $N\Delta$ interactions represent the first resonance contribution to the single pion production in $NN$ collisions~\cite{Engel:1996ic,Machleidt:2011zz} and hence have important consequences on the properties of dense nuclear matter~\cite{Drago:2014oja,Cai:2015hya,Sedrakian:2022ata,Ribes:2019kno,Harris:2024ssp} and transport phenomena in heavy ion collisions~\cite{SMASH:2016zqf}. These couplings are also of importance in understanding the electromagnetic and gravitational $N\to\Delta$ transition form factors~\cite{Jones:1972ky,Pascalutsa:1999zz,Pascalutsa:2006up,Kim:2024hhd,Alharazin:2023zzc} which can be accessed through hard exclusive processes at the upcoming Electron-Ion Collider~\cite{AbdulKhalek:2021gbh}. The $N\Delta$ couplings are also important for the search of exotic states like dibaryons~\cite{Mulders:1980vx,Li:2000cb,Valcarce:2005em,Lu:2020qme}. A key challenge in extracting these couplings from data is that a majority of them correspond to subthreshold decays (\eg, $\Delta\to \rho N$) and hence are not readily accessible in scattering or production experiments.

High-energy scattering offers complementary access to these couplings. The near-forward region is dominated by $t$-channel Regge exchanges, each summing an infinite tower of states of increasing spin, given naturality, and  signature~\cite{Collins:1977jy,JointPhysicsAnalysisCenter:2024kck,Winney:2025tla}. A high-energy scattering process therefore carries information about the couplings of all the states on a given Regge trajectory to the external states. In principle, it is possible to extract these $t$-channel residues using the crossing relations between $s$- and $t$-channel amplitudes and analyticity of the $t$-channel helicity amplitudes. In practice, such a program requires experimental data which constrains both the relative size and the complex phase of the different helicity amplitudes. This information can be obtained from polarized observables, which until recently have not been available.

Given that Regge theory does not constrain these couplings beyond analyticity, in the absence of polarized data, they have typically been modeled using quark models~\cite{Arenhovel:1975vf} with effective Lagrangian approaches~\cite{Gotsman:1969rsi,Barbour:1974nr,Clark:1977ce,Nam:2011np,Yu:2016jfi,JointPhysicsAnalysisCenter:2017del}. The recent availability of high-precision data of the spin density matrix elements (SDMEs) of the polarized photoproduction of $\pi\Delta$ from GlueX~\cite{GlueX:2024dbr} has thus opened a new era in hadron spectroscopy where the residues of subthreshold couplings can also be extracted directly from scattering data with good precision.\par

In this Letter, we present the first direct extraction of $t$-channel Regge residues from data, from which the couplings can be extracted. Starting from a fit of the $s$-channel amplitudes, we construct the $t$-channel partial waves and isolate the residues of the exchanges of the lowest-lying states in each of the Regge trajectories. We apply the procedure to high-energy $\pi\Delta$ photoproduction, using the  GlueX data on SDMEs~\cite{GlueX:2024dbr} and the SLAC data for the cross section~\cite{Boyarski:1968dw}. For pion exchange, we extract the $\pi N\Delta$ coupling constant from the residue and find good agreement with that obtained from the decay width~\cite{ParticleDataGroup:2024cfk}. For the first time, we determine the complete set of $N\Delta$ couplings to $\rho$, $b_1$, and $a_2$ corresponding to all possible spin-orbit combinations. Our work forges a pathway for robust extractions of such couplings for other systems of resonances whenever data become available.

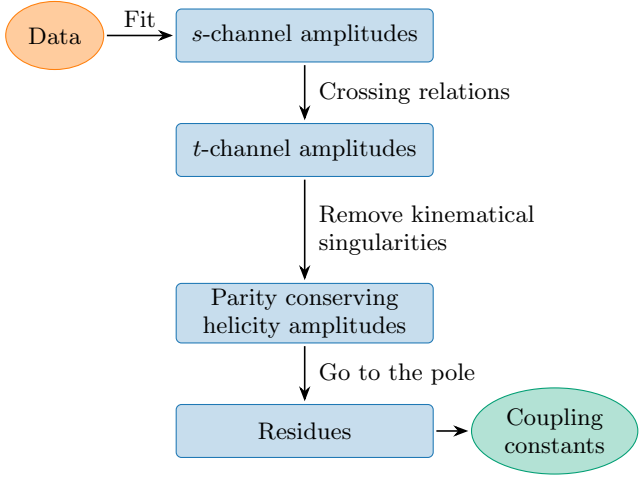
\begin{figure}[t]
\begin{tikzpicture}[
    amp/.style={draw=jpac-blue, fill=jpac-blue!25, rounded corners=2pt, minimum width=34mm, minimum height=7mm, align=center, font=\small, inner sep=2pt},
    data/.style={draw=jpac-orange, fill=jpac-orange!40, ellipse, minimum width=10mm, minimum height=9mm, font=\small},
    output/.style={draw=jpac-green, fill=jpac-green!30, ellipse, minimum width=22mm, minimum height=11mm, align=center, font=\small},
    arr/.style={-{Stealth[length=2mm]}, semithick, shorten >=1pt, shorten <=1pt},
    proclbl/.style={font=\small, align=left, inner sep=3pt}
]
    \node[amp] (samp) {$s$-channel amplitudes};
    \node[amp, below=8mm of samp]  (tamp) {$t$-channel amplitudes};
    \node[amp, below=14mm of tamp] (ksf)  {Parity conserving\\ helicity amplitudes};
    \node[amp, below=8mm of ksf]   (res)  {Residues};
    \node[data] (data) at ($(samp.west)-(16mm,0)$) {Data};
    \node[output] (coup) at ($(res.east) +(16mm,0)$) {Coupling\\ constants};
    \draw[arr] (samp) -- (tamp)
        node[proclbl, midway, right=2pt] {Crossing relations};
    \draw[arr] (tamp) -- (ksf)
        node[proclbl, midway, right=2pt] {Remove kinematical\\ singularities};
    \draw[arr] (ksf) -- (res)
        node[proclbl, midway, right=2pt] {Go to the pole};
    \draw[arr] (data) -- (samp)
        node[proclbl, midway, above] {Fit};
    \draw[arr] (res) -- (coup);
\end{tikzpicture}
    \caption{Flowchart of the methodology employed to obtain the residues. Also shown is a use-case where we extract the coupling constants of an effective Lagrangian. See text for details.}
    \label{fig:flowGM}
\end{figure}

\mytitle{Residues and couplings} 
Since the physical process ($\gamma p \to \pi \Delta$) is in the $s$-channel frame, it is natural to choose it to write the amplitude. Prior works have indeed used definitions based on $s$-channel Regge residues without crossing to the $t$-channel frame~\cite{Irving:1977ea,Ebel:1970wi,Ebel:1971cd,Pilkuhn:1973wq,Nagels:1976mc,Nagels:1979xh} and therefore they cannot yet be identified with intermediate states of the $t$-channel, i.e. $\gamma \bar{\pi} \to \bar{p}\Delta$, reaction.\footnote{To align with the standard notations in the literature, throughout the paper we refer to nucleon ($N$) or proton ($p$) interchangeably. Consequently, the $\Delta$ and $\pi$ are always $\Delta^{++}$ and $\pi^-$.}
 
The rigorous definition of the residues is obtained from the $t$-channel helicity amplitudes, where  we can naturally relate the exchanges to physical states. At large $s$ and in the vicinity of a pole of a state with spin $J$, mass $m_J$, and naturality $\eta$, the relation between the $t$-channel partial wave amplitude and its residue reads
\begin{align}
    T^{(t)J,\eta}_{\helgammat\helprotont\helDelt} (t,s) &\propto 
    \mathcal{R}^{J\eta}_{\helgammat\helprotont\helDelt} 
    \frac{s^J}{t - m_J^2} + \text{finite terms}
    \ , \label{eq:resdef1}
\end{align}
where we have omitted the known kinematic functions. The subindices represent the helicities corresponding to the $t$-channel process. The residue $\mathcal{R}^{J\eta}_{\helgammat\helprotont\helDelt}$ encodes the dynamics of the corresponding pole exchange process and can be matched to a Lagrangian in order to extract the various couplings. This definition motivates the strategy used in the following analysis, and is shown schematically in Fig.~\ref{fig:flowGM}.

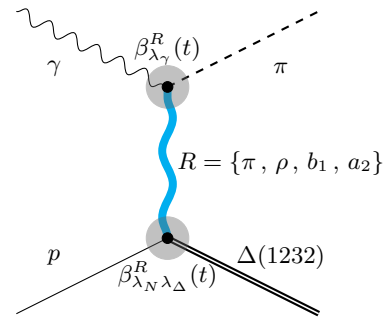
\begin{figure}[b]
    \centering
    \begin{tikzpicture}
        \draw (0,0) -- (2,1);
        \draw[double,thick] (2,1) -- (4,0);
        \draw[decorate,decoration={snake}] (0,4) -- (2,3);
        \draw[dashed,thick] (2,3) -- (4,4);
        \draw[decorate,color=cyan,line width=3pt,decoration={snake,amplitude=2,segment length=25}] (2,1) -- (2,3);
        \filldraw[black!60,opacity=0.4] (2,1) circle (8pt);
        \filldraw[black!60,opacity=0.4] (2,3) circle (8pt);
        \filldraw[black] (2,1) circle (2pt);
        \filldraw[black] (2,3) circle (2pt);
        \node at (0.5,0.75) {$p$};
        \node at (0.5,3.25) {$\gamma$};
        \node at (3.5,0.75) {$\DelBar$};
        \node at (3.5,3.25) {$\pi$};
        \node at (3.5,2) {$R = \{\pi\,,\,\rho\,,\,b_1\,,\,a_2\}$};
        \node at (2,3.5) {$\beta^R_{\helgamma}(t)$};
        \node at (2,0.5) {$\beta^R_{\helproton\helDel}(t)$};
    \end{tikzpicture}
    \caption{The Regge exchange diagram describing the $\pi\Delta$ photoproduction. The model encodes factorized Regge couplings in the $s$-channel~\cite{Shastry:2026vlf}.}\label{fig:feyndia}
\end{figure}

\begin{figure*}[t]
    \centering
    \centering
    \includegraphics[width=0.3\linewidth]{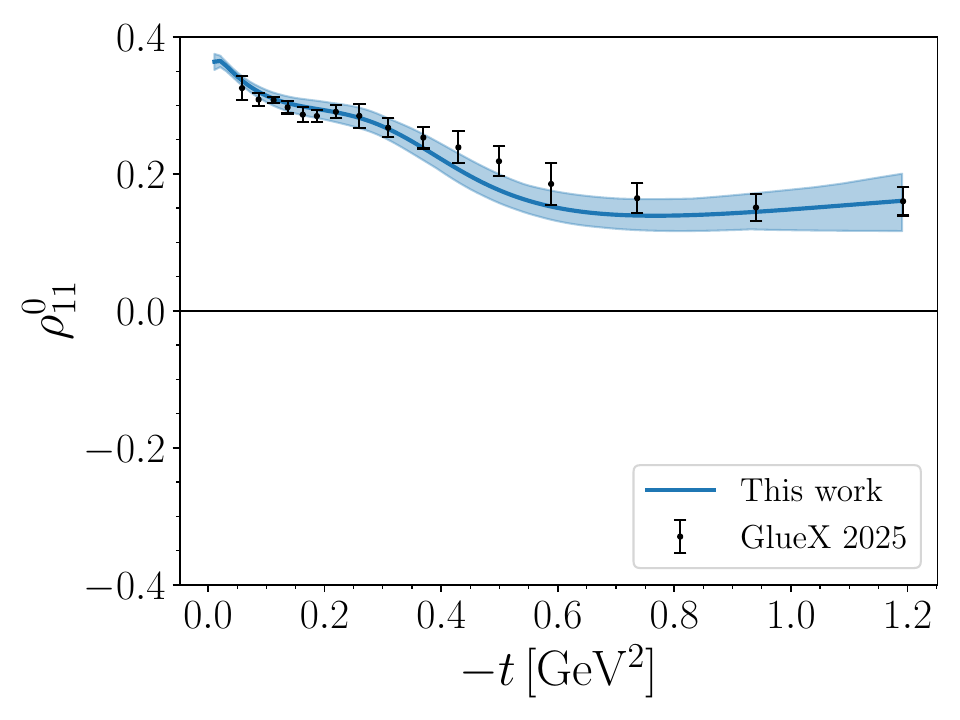}
    \includegraphics[width=0.3\linewidth]{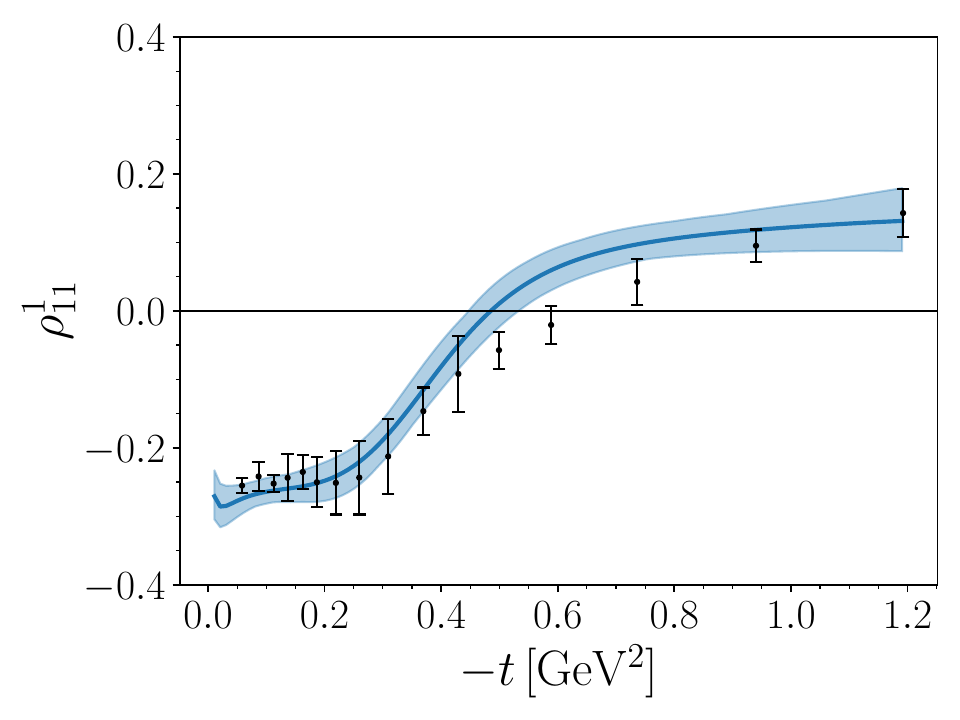}~\includegraphics[width=0.3\linewidth]{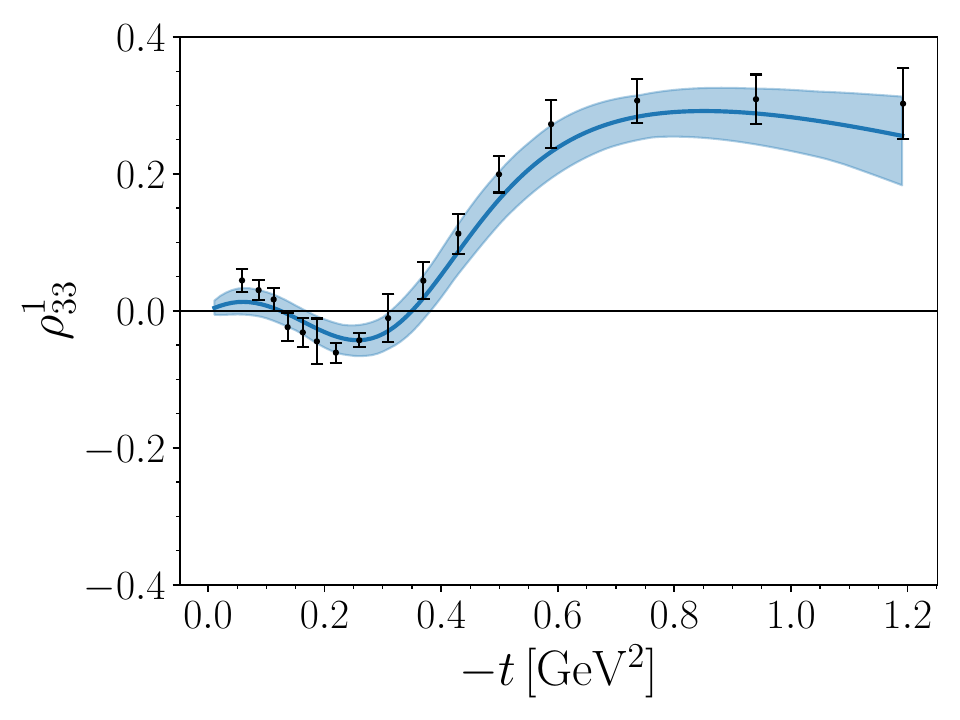}
    \caption{SDMEs in the Gottfried-Jackson frame from the present work compared to the GlueX data~\cite{GlueX:2024erj}. We refer the reader to the companion paper for the details of the calculations~\cite{Shastry:2026vlf}.}
    \label{fig:SDMEs}
\end{figure*}

\mytitle{Regge exchange amplitude}
We model the $s$-channel amplitudes as a sum of $\pi$, $\rho$, $b_1$, and $a_2$ Regge exchanges (\cf Fig.~\ref{fig:feyndia}).
We use a factorized form of the vertices~\cite{Cohen-Tannoudji:1968eoa}, except for subleading corrections related to final state interactions in the case of unnatural exchanges~\cite{Williams:1970rg,JointPhysicsAnalysisCenter:2017del}. The vertex involving the external $\gamma\pi$ can be modeled using Lagrangians, with the couplings fixed by the measured radiative decay widths. For the $p\Delta$ vertex we use generic polynomials in $t$ for each independent helicity combination supplemented with exponential form factors to account for the observed behavior of the cross section~\cite{Shastry:2026vlf}. The coefficients are extracted by a simultaneous fit to the SDME data from GlueX~\cite{GlueX:2024dbr} in the Gottfried-Jackson frame, and to the cross-section data from SLAC~\cite{Boyarski:1968dw}, which together constrain both the magnitude and sign of the couplings. The model has 20 free parameters and the fit gives $\chi^2/\text{dof} \simeq 1.1$. A complete description of the amplitude, fit procedure, and the values of the fitted parameters can be found in the companion paper~\cite{Shastry:2026vlf}. The diagonal SDMEs are shown in Fig.~\ref{fig:SDMEs}. An important feature visible in the SDMEs is the approximate symmetry between \mbox{$\rho^0_{\helDelt,\helDelt}$} and \mbox{$\rho^1_{\helDelt,\helDelt}$} which arises from the factorization of the vertices and the dominance of the exchanges of a particular naturality in a given region of $t$: unnatural exchanges dominate at small $|t|$, while the natural ones dominate at large $|t|$.

\mytitle{Crossing relations and residue extraction} 
We cross the amplitude from the $s$- to the $t$-channel using the relations in Ref.~\cite{Collins:1971ff}. To access the residues, we construct partial waves of definite spin and parity which then need to be continued to timelike values of $t$. However, these present a number of kinematical singularities~\cite{JPAC:2018dfc,Collins:1971ff,Jackson:1968rfn,Henyey:1968azc}, which must be removed in order to get the residues in the correct Riemann sheet. The prescription to do so is given in the companion paper~\cite{Shastry:2026vlf}. To validate the approach, we extract the $\pi N\Delta$ coupling from the $\pi$-exchange amplitude as \mbox{$f_{\pi N\Delta} = -2.18 \pm 0.08$} which gives the $\Delta(1232)$ decay width \mbox{$\Gamma_{\Delta\to p\pi} = 129 \pm 9\text{ MeV}$}, consistent with the value listed in the PDG~\cite{ParticleDataGroup:2024cfk}.

Now that we have validated the procedure, we can move to extract the couplings of the mesons corresponding to the kinematically forbidden decays of the $\Delta$. Unlike the case of the pion, for higher spin exchanges we can construct multiple independent spin-orbit combinations, of which the previous lack of polarized data has prevented a full extraction. We tabulate the complete set of these couplings in Table~\ref{tab:couplings}. The particular hierarchy in the values of the coupling constants is an artifact of the choice of basis and the choice of normalization of the coupling constants.

In particular, the $\rho N\Delta$ vertex has been used in the past typically assuming \mbox{$g^{(1)}_{\rho N\Delta} = g^{(3)}_{\rho N\Delta} = 0$}~\cite{Engel:1996ic,Nam:2011np,Janssen:1996kx,Ronchen:2012eg} while $g^{(2)}_{\rho N\Delta}$ was constrained using a quark model~\cite{Arenhovel:1975vf}. Since we ease this restriction, our values differ significantly.

\begin{table}[b]
    \caption{Couplings of the various mesons to $N\Delta$ extracted from the residues. Our approach allows for the extraction of not only the magnitudes of the coupling constants but also their relative signs. We divide out the coupling for the upper vertices, calculated from the radiative widths~\cite{Shastry:2026vlf}.}
    \label{tab:couplings}
    \centering
    \begin{tabular}{|c|c|c|c|}
         \hline
         \backslashbox{$g^{(i)}_{R N \Delta}$}{$R\quad$} & $\rho$ & $b_1$ & $a_2$\\\hline   
         $g^{(1)}_{R N\Delta}$ & $33 \pm 10$ &  $-1.1 \pm 1.9$ & $-69 \pm 27$\\
         $g^{(2)}_{R N\Delta}$ & $-522 \pm 82$ & $-199 \pm 98$ & $627 \pm 151$\\
         $g^{(3)}_{R N\Delta}$ & $53 \pm 19$ & $66 \pm 37$  & $53 \pm 40$\\
         $g^{(4)}_{R N\Delta}$ & --- & --- & $6.6 \pm 5.0$ \\\hline
    \end{tabular}
\end{table}

The couplings to $b_1$ and $a_2$ are also shown in Table~\ref{tab:couplings}, albeit with larger uncertainties. To the best of our knowledge, these couplings have not been studied in the literature, neither estimated from quark model nor extracted from any data.

\mytitle{Summary and conclusions} We have shown how to use Regge theory to systematically extract the $t$-channel residues from high-energy data, and applied it to $\pi\Delta$ photoproduction. We first fit the $s$-channel Regge couplings, parametrized as polynomials in $t$, to the available experimental data {\it viz.,} the recent SDME measurements from GlueX and the cross section from SLAC. Crossing the helicity amplitudes to the $t$-channel with careful consideration of kinematical singularities gives direct access to the residues of the various meson exchanges. Because a residue is a property of the $S$-matrix, it provides a rigorous characterization of the interaction, from which the coupling constants follow. We have validated the method on pion exchange and provide, for the first time, the complete set of $N\Delta$ couplings to $\rho,\, b_1,$ and $ a_2$, which are not accessible from $\Delta$ decays. The $\rho N\Delta$ couplings, in particular, differ significantly from the model values used in some of the earlier works on single pion production in $NN$ collisions~\cite{Engel:1996ic} or in the predictions of dibaryons~\cite{Mulders:1980vx,Li:2000cb,Valcarce:2005em,Lu:2020qme}.

This method establishes high-energy production reactions as a quantitative, robust source for resonance couplings. Importantly, these include the couplings to states that are kinematically forbidden in the decay and are therefore inaccessible to conventional low-energy partial wave analyses. We emphasize the importance of high-statistics measurements of polarization observables, such as SDMEs, for such studies.

\begin{acknowledgments}
\mytitle{Acknowledgments} The authors thank F.~Afzal, A.~Schertz, M.~Shepherd, and J.~Stevens from the GlueX collaboration for useful discussions. This work was supported by the U.S.~Department of Energy contract \mbox{DE-AC05-06OR23177}, under which Jefferson Science Associates, LLC operates Jefferson Lab, by U.S.~Department of Energy Grant Nos.~\mbox{DE-FG02-87ER40365}, and \mbox{DE-SC0011090}, and it contributes to the aims of the U.S.~Department of Energy \mbox{ExoHad} Topical Collaboration, contract \mbox{DE-SC0023598}. {\L}B was partially financed by the Faculty of Physics and Applied Computer Science AGH~UST statutory tasks within subsidy of Ministry of Science and Higher Education. GM and VM have been supported by projects \mbox{CEX2024-001451-M} (Unidad de Excelencia ``María de Maeztu''), \mbox{PID2023-147112NB-C21}, all financed by \mbox{MICIU/AEI/10.13039/501100011033/} and FEDER, UE. GM is a Beatriu de Pin\'os Fellow. VM is a Professor Serra H\'unter. VM acknowledges support from \mbox{CNS2022-136085}. RJP acknowledges support by the Simons Foundation award Simons Collaboration on Confinement and \mbox{QCD StringsMPS-QCD-00994314} and by MIT. 
\end{acknowledgments}

\bibliography{Ref}
\bibliographystyle{apsrev4-2}
\end{document}